\documentclass[MSNbibl,nameyear,dvips]{arxstspdf}
\usepackage{subenv}
\usepackage{graphicx}
\usepackage{flushend}
\usepackage{stfloats}
\volume{27}
\issue{3}
\pubyear{2012}
\firstpage{395}
\lastpage{411}
\doi{10.1214/12-STS390} 

\makeatletter
\newcommand{\citecs}[1]{\citeauthor{#1}, \citeyear{#1}}
\makeatother

\begin{document}
\begin{frontmatter}

\title{Analysis of 1:1 Matched Cohort Studies and Twin Studies, with Binary Exposures and Binary Outcomes}

\runtitle{Matched cohort studies}

\begin{aug}
\author[a]{\fnms{Arvid} \snm{Sj\"olander}\corref{}\ead[label=e1]{arvid.sjolander@ki.se}},
\author[b]{\fnms{Anna L. V.} \snm{Johansson}\ead[label=e2]{anna.johansson@ki.se}},
\author[c]{\fnms{Cecilia} \snm{Lundholm}\ead[label=e3]{cecilia.lundholm@ki.se}},
\author[d]{\fnms{Daniel} \snm{Altman}\ead[label=e4]{daniel.altman@ki.se}},
\author[e]{\fnms{Catarina}~\snm{Almqvist}\ead[label=e5]{catarina.almqvist@ki.se}}
\and
\author[f]{\fnms{Yudi} \snm{Pawitan}\ead[label=e6]{yudi.pawitan@ki.se}}

\runauthor{A. Sj\"olander et al.}

\affiliation{Karolinska Institutet}

\address[a]{Arvid Sj\"olander is Ph.D. Student, Department of Medical Epidemiology
and Biostatistics, Karolinska Institutet, Solna, Sweden \printead{e1}.}
\address[b]{Anna L. V. Johansson is Statistician, Department of Medical Epidemiology
and Biostatistics, Karolinska Institutet, Solna, Sweden \printead{e2}.}
\address[c]{Cecilia Lundholm is Statistician, Department of Medical Epidemiology
and Biostatistics, Karolinska Institutet, Solna, Sweden \printead{e3}.}
\address[d]{Daniel Altman is Associate Professor, Department of Medical Epidemiology
and Biostatistics, Karolinska Institutet, Solna, Sweden, and Associate Professor,
Division of Obstetrics and Gynecology, Department of Clinical Sciences, Danderyd Hospital,
Karolinska Institutet, Stockholm, Sweden \printead{e4}.}
\address[e]{Catarina Almqvist is Associate Professor, Department of Medical Epidemiology
and Biostatistics, Karolinska Institutet, Solna, Sweden, and Associate Professor,
Astrid Lindgren Children's Hospital and Department of Woman and Child Health,
Karolinska Institutet, Stockholm, Sweden \printead{e5}.}
\address[f]{Yudi Pawitan is Professor, Department of Medical Epidemiology
and Biostatistics, Karolinska Institutet, Solna, Sweden \printead{e6}.}

\end{aug}

\begin{abstract}
To improve confounder adjustments, observational studies are often
matched on potential confounders. While matched case-control studies
are common and well covered in the literature, our focus here is on
matched cohort studies, which are less common and sparsely discussed in
the literature. Matched data also arise naturally in twin studies, as a
cohort of exposure--discordant twins can be viewed as being matched on
a large number of potential confounders. The analysis of twin studies
will be given special attention. We give an overview of various
analysis methods for matched cohort studies with binary exposures and
binary outcomes. In particular, our aim is to answer the following
questions: (1) What are the target parameters in the common analysis
methods? (2) What are the underlying assumptions in these methods? (3)
How do the methods compare in terms of statistical power?
\end{abstract}


\end{frontmatter}

\section{Introduction}\label{sec:introduction}

A common goal of epidemiological research is to estimate the causal
effect of a particular exposure on a particular outcome. The common
tool is an observational study, utilizing, for example, hospital data,
cohort data or health register data. In observational studies, the
exposure-outcome association is invariably confounded by factors that
induce spurious (i.e., noncausal) associations. For example, age may
confound an exposure-outcome association if older people are more often
exposed and more likely to develop the outcome. Without adjustment for
age, that is, if the confounding influence by age is not accounted for
in the analysis, there may be an association of exposure and outcome,
even in the absence of a causal effect. Hence, the exposure-outcome
association cannot, in general, be given a causal interpretation,
unless all confounders are properly adjusted for.

There are several strategies to adjust for potential confounders in
the analysis, for example, stratification or regression modeling.
Essentially, these methods solve the problem of confounding by
comparing the exposed and unexposed within levels of the
confounders, thus balancing the confounders across levels of the
exposure and comparing ``like with like.'' If there is a strong
association between the confound\-ers and the exposure, or between the
confounders and the outcome, these strategies are often inefficient.
In particular, some strata may contain few exposed subjects or few
cases (i.e., subjects that developed the outcome); the lack of
balance may lead to unstable estimates for these
strata.

One common method to increase the efficiency is to  match the study on
potential confounders. For example, matched case-control studies are
constructed so that for each case, a fixed number of controls are
selected, having the same confounder levels as the case. When each case
is matched to one control, we say that the study is 1:1 matched. In
case-control studies, matching forces the ratio of cases to controls to
be constant across all strata of the matched factors, which implies
that the association between the confounders and the outcome is broken.
Matched case-control studies are commonplace, and well covered in the
literature (e.g., \citecs{Breslow}; \citecs{Jewell};
\citecs{Woodward}). A matched cohort study can be constructed in a
similar fashion; for each exposed subject, a fixed number of unexposed
subjects are selected, having the same confounder levels as the
exposed. In cohort studies, matching forces the ratio of exposed to
unexposed to be constant across all strata of the matched factors,
which implies that the association between the confounders and the
exposure is broken. Matched cohort studies are relatively rare, and the
literature is sparse and typically rather brief (e.g.,
Cummings et al., \citeyear{Cummings}). The reason, we believe, is mainly due to available
data sources. Matched cohort studies are suitable for situations where
a researcher has access to large population data sources with exposure
information.

Matched data also arise naturally in  twin studies. By nature, a large
number of potential confounders are shared (i.e., having constant
levels) within each twin pair, for example, genetic factors, maternal
uterine environment, gestational age, etc. It follows that a cohort
of\vadjust{\goodbreak}
exposure--discordant twin pairs (i.e., pairs in which one of the twins
is exposed, and the other twin is unexposed) can be viewed as being 1:1
matched on all shared confounders. In such a cohort there is no
association between the shared confounders and the exposure. An
attractive feature of twin studies is that the shared confounders often
include factors which are normally very difficult to match on, or even
to measure. For example, monozygotic twins have identical genes and can
thus can be viewed~as being matched on the whole genome. However,
a~twin study is not simply a special case of a regular 1:1 matched cohort
study; whereas the latter only contains exposure--discordant pairs, the
former also contains pairs which are concordant in the exposure.~Be\-cause
of their unique and attractive properties, twin studies will be
given special attention in this paper.\looseness=-1

The aim of this paper is to give a detailed overview of different
analysis methods for matched cohort studies with binary exposures and binary outcomes. In
particular, our aim is to answer the following questions: (1) What are the target parameters in the common analysis methods? (2) What are the underlying assumptions in these methods? (3) How do the methods compare in terms
of statistical power?

We illustrate the methods with two examples. The first
example is a register-based study on the effect of hysterectomy on
the risk for cardiovascular disease (CVD) in Swedish women
\citep{Ingelsson}. The study is matched on birth year, year of hysterectomy and county of residence at year of
hysterectomy, so that for each hysterectomized woman (exposed),
three nonhysterectomized women at same age and year were selected
from the general population. The second study is a population-based
twin study of the association between fetal growth and childhood
asthma \citep{Ortqvist}.

The paper is organized as follows. In Section~\ref{sec:margcondstd} we
review the concepts of marginalization, conditioning and
standardization. In Section~\ref{sec:matched} we define a matched
cohort study. In Section~\ref{sec:analysis} we describe the most common
analysis methods for matched cohorts. These methods can also be used to
analyze the exposure--discordant pairs in twin studies. In Section~\ref{sec:twins} we demonstrate how these methods can be adapted for
inclusion of the exposure--concordant pairs in twin studies as well. In
Section~\ref{sec:sim} we carry out a simulation study. In Section~\ref{sec:examples} we provide the two illustrating examples. We will
restrict our attention to 1:1 matching, and we will not consider
additional covariate adjustments. Extensions to other matching schemes
and adjustments for additional covariates are discussed in Section~\ref{sec:discussion}.

\section{Marginalization, Conditioning and Standardization}
\label{sec:margcondstd}

We first establish the notations and briefly review the concepts of
marginalization, conditioning  and standardization, which are crucial
for the understanding of matching and confounder adjustment. More
thorough discussions can be found in standard epidemiological textbooks
(e.g., Rothman et~al.,\break \citeyear{Rothman}). Let $X$ denote the binary exposure of
interest (0/1), let $Y$ denote the binary outcome of interest (0/1) and
let $Z$ denote a set of potential confounders for the association
between $X$ and $Y$. We use $\operatorname{Pr}(\cdot)$ generically for
both probabilities (population proportions) and densities, and we use
$E(\cdot)$ for expected value (population average). We use $V_1\perp
V_2|V_3$ as shorthand for ``$V_1$ and $V_2$ conditionally independent,
given $V_3$.'' We use (log) odds ratios to quantify the $X$--$Y$
association. Other possible options would be risk differences or risk
ratios. There are two reasons for focusing on odds ratios. First,
regression models for odds ratios can be conveniently fitted without
restrictions; see Section~\ref{sec:rtrt}. Second, in applied scenarios,
it is often desirable to make results comparable with case control
studies, in which only odds ratios are estimable.

An unadjusted analysis targets the marginal\break (over~$Z$)  association
between $X$ and $Y$, for example, through the marginal odds ratio
\[
\mathit{OR}_m=\frac{\operatorname{Pr}(Y=1|X=1)\operatorname{Pr}(Y=0|X=0)}{\operatorname{Pr}(Y=0|X=1)\operatorname{Pr}(Y=1|X=0)}.
\]
We define $\psi_m=\log(\mathit{OR}_m)$. In the presence of confounders
$Z$, $\mathit{OR}_m$ fails to have a causal interpretation. In
particular, it may differ from 1 in the absence of a causal effect.

The influence of $Z$ can be eliminated by conditioning on $Z$, as in the conditional odds ratio
\[
\mathit{OR}_c(Z)=\frac{\operatorname{Pr}(Y=1|X=1,Z)\operatorname{Pr}(Y=0|X=0,Z)}{\operatorname{Pr}(Y=0|X=1,Z)\operatorname{Pr}(Y=1|X=0,Z)}.
\]
The conditional odds ratio $\mathit{OR}_c(Z)$ depends, in general, on
$Z$. If $Z$ is the only confounder for the $X$--$Y$ association, then
$\mathit{OR}_c(Z)$ can be interpreted as the conditional causal effect
of $X$ on $Y$, given $Z$, on the odds ratio scale. If there are
additional confounders, then  $\mathit{OR}_c(Z)$ has no causal
interpretation.

$\mathit{OR}_c(Z)$ is a subpopulation (i.e., $Z$-specific) effect. The
effect for the whole population can be obtained through
standardization. The standardized probability of $Y=1$ given $X=x$, is
given by
\begin{equation}
\label{eq:std} E_Z\{\operatorname{Pr}(Y=1|X=x,Z)\},
\end{equation}
where we have used subindex $Z$ to highlight that the expectation is
taken over the marginal distribution $\operatorname{Pr}(Z)$. We
emphasize that the expression in (\ref{eq:std}) is not, in general,
equal to
$E_{Z|X=x}\{\operatorname{Pr}(Y=1|X=x,Z)|X=x\}=\operatorname{Pr}(Y=1|X=x)$,
which is the marginal (unadjusted) probability of $Y=1$, given $X=x$.
If $Z$ is the only confounder, then
$E_Z\{\operatorname{Pr}(Y=1|X=x,Z)\}$ can be interpreted as the
hypothetical (counterfactual) probability of $Y=1$, had everybody
attained level $X=x$ in the source population \citep{Hernan}.
$\operatorname{Pr}(Y=1|X=x,Z)$ can be standardized to any proper
distribution $\operatorname{Pr}^{*}(Z)$, not necessarily equal to
$\operatorname{Pr}(Z)$. We let $E^*_Z(V)$ denote the expected value of
$V$, where the expectation is taken over $\operatorname{Pr}^{*}(Z)$. If
$Z$ is the only confounder, then
$E^*_Z\{\operatorname{Pr}(Y=1|X=x,Z)\}$ can be interpreted as the
hypothetical (counterfactual) probability of $Y=1$, had everybody
attained level $X=x$ in the fictitious population where $Z$ follows the
distribution $\operatorname{Pr}^{*}(Z)$. A standardized odds ratio is
constructed as
\begin{eqnarray*}
&&\hspace*{-5pt}\mathit{OR}_s
\\
&&\hspace*{-5pt}\quad=\frac{E\{\operatorname{Pr}(Y=1|X=1,Z)\}E\{\operatorname{Pr}(Y=0|X=0,Z)\}}{E\{\operatorname{Pr}(Y=0|X=1,Z)\}E\{\operatorname{Pr}(Y=1|X=0,Z)\}}.
\end{eqnarray*}
We define $\psi_s=\log(\mathit{OR}_s)$. In ($\ref{eq:std}$),
$\operatorname{Pr}(Y=1|X=x,Z)$ is standardized to
$\operatorname{Pr}(Z)$, that is, the distribution of~$Z$ in the source
population. In order to keep the notation simple, we use
$\mathit{OR}_s$ and $\psi_s$, even if $\operatorname{Pr}(Z)$ is
replaced by $\operatorname{Pr}^*(Z)$, and we let it be clear from the
context which distribution of $Z$ these parameters are standardized to.
If $Z$ is the only confounder, then $\mathit{OR}_s$ can be interpreted
as the causal effect of~$X$ on~$Y$ in the source/fictitious population,
on the odds ratio scale. We emphasize that although the numerical
values of $\mathit{OR}_s$ and $\psi_s$ may depend heavily on which
distribution of $Z$ they are standardized to, they are always, by
construction, adjusted for $Z$.

In general, there is no ordering in the magnitudes of
$\mathit{OR}_c(Z)$, and $\mathit{OR}_s$. An interesting special case
occurs when $\mathit{OR}_c(Z)$ is constant across levels of $Z$, that
is,
\begin{equation}
\label{eq:const} \log\{\mathit{OR}_c(Z)\}=\psi_c.
\end{equation}
It can be shown \citep{Neuhaus4} that $|\psi_c|\geq|\psi_s|$.

\begin{table*}[b]
\vspace*{3pt}
\tabcolsep=0pt
\caption{Crude summary of matched 1:1 cohort data}\label{tab:tab1}
\begin{tabular*}{290pt}{@{\extracolsep{\fill}}lccc@{}}
\hline
&\multicolumn{2}{@{}c}{\textbf{Unexposed pair member ($\bolds{X=0}$)}}& \textbf{Totals}\\
\ccline{2-3}
& \textbf{Event ($\bolds{Y=1}$)}             & \textbf{\hspace*{3pt}No event ($\bolds{Y=0}$)\hspace*{-4pt}} &\\
\hline
Exposed pair member ($X=1$) &           &           &\\
\quad Event ($Y=1$)         & $T$       & $U$       & $T+U$\\
\quad No event ($Y=0$)      & $V$       & $W$       & $V+W$\\
\quad Totals                & $T+V$     & $U+W$     & $n$\\
\hline
\end{tabular*}
\end{table*}

\begin{figure}

\includegraphics{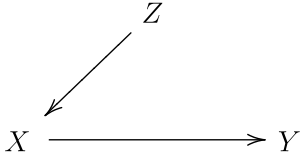}

\caption{A causal structure for which $Y\perp Z|X$.}
\label{fig:d2}
\end{figure}

\begin{figure}

\includegraphics{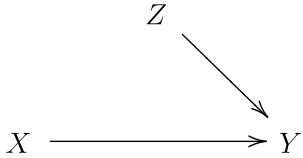}

\caption{A causal structure for which $X\perp Z$.}
\label{fig:d1}
\end{figure}

In general, there is no ordering in the magnitudes of $\mathit{OR}_m$
and $\mathit{OR}_c(Z)$,  or of $\mathit{OR}_m$ and $\mathit{OR}_s$;
confounding by $Z$ can both inflate or deflate the association between
$X$ and $Y$. There are a few special cases though. If $Y\perp Z|X$, then
$\operatorname{Pr}(Y=1|X,Z)=\operatorname{Pr}(Y=1|X)$ which
implies
that $\mathit{OR}_m=\mathit{OR}_c(Z)=\mathit{OR}_s$ for all\vadjust{\goodbreak} $Z$ and all
standardization distributions $\operatorname{Pr}^*(Z)$. This would
happen if the true causal structure between $X$, $Y$ and $Z$ is as in
Figure \ref{fig:d2}. If $X\perp Z$, then
$\operatorname{Pr}(Z|X)=\operatorname{Pr}(Z)$ which implies that
$\mathit{OR}_m=\mathit{OR}_s$ for the particular distribution
$\operatorname{Pr}(Z)$, that is, the distribution of $Z$ in the source
population. This would happen if the true causal structure is as in
Figure \ref{fig:d1}.

We note that in Figures \ref{fig:d2} and \ref{fig:d1}, $Z$ is not a
confounder, and $\mathit{OR}_m$ can be given a causal interpretation.
Thus, for these scenarios, adjusting for $Z$ is not necessary for
causal inference. We further note that the structure in Figure
\ref{fig:d1} does not render $\mathit{OR}_m$ equal to
$\mathit{OR}_c(Z)$, even if $\mathit{OR}_c(Z)$ is constant across
levels of $Z$. This is a consequence of the noncollapsibility of the
odds ratio. For a more thorough discussion on (non)collapsibility and
the special properties of odds ratios, we refer the reader to
\citet{Greenland}.\looseness=-1

\section{Matched Cohort Studies}
\label{sec:matched}

\subsection{Design}

A cohort study that is 1:1 matched on $Z$ consists of $n$ pairs of
observations, each pair consisting of one exposed subject ($X=1$) and
one unexposed subject ($X=0$). The pairs are constructed so that the two
subjects within each pair have the same level of confounder $Z$;
that
is, $Z$ may vary between pairs, but not within pairs. Thus, $Z$ is
equally distributed among exposed and unexposed in the matched cohort.
The outcome $Y$ is assumed to be recorded for each subject. Ignoring
$Z$, the paired data can be conveniently represented as in Table
\ref{tab:tab1}. In practice, 1:1 matched pairs are typically
constructed by first drawing an exposed person from the whole
population, then drawing an unexposed person with an equal or similar
level of confounder $Z$; we refer to this sampling scheme as \emph{exposure-driven}
matching.

We note that in twin studies $Z$ is not directly observed, but should
be interpreted as all the unobserved factors that are common within a
twin pair.

\subsection{Likelihood Construction}

Before discussing the various analysis methods, we construct the
likelihood for the observed data. Let $Z_i$ denote the common value of
$Z$ for pair $i$,  $i\in\{1,2,\ldots,n\}$. Let $Y^0_i$ and $Y^1_i$
denote the outcome $Y$ for the unexposed ($X=0$) and the exposed
($X=1$) subject in pair $i$, respectively. The matched data consists of
$n$ i.i.d. observations ($Y^0_i,Y^1_i,Z_i$). We suppress the index $i$
when not needed, so that $Y^x$ denotes $Y$ for the subject with $X=x$,
$x\in (0,1)$, within an arbitrary pair. We use
$\operatorname{Pr}(Y=y,X=x,Z=z)$ to denote the population probability
of ($Y=y,{X=x},\break Z=z$), and we will use
$\operatorname{Pr}^*(Y^0=y^0,Y^1=y^1,Z=z)$ to denote the probability
for ($Y^0=y^0,Y^1=y^1,{Z=z}$) induced by the matched sampling scheme.
Under exposure-driven matching, the design implies that
\begin{subequation}
\begin{equation}
\label{eq:faca}
\operatorname{Pr}^*(Y^x=y^x|Z)=\operatorname{Pr}(Y=y^x|X=x,Z)
\end{equation}
and \begin{eqnarray}\label{eq:facb}
&&\operatorname{Pr}^*(Y^0=y^0,Y^1=y^1|Z)\nonumber
\\[-8pt]\\[-8pt]
&&\quad=\operatorname{Pr}^*(Y^0=y^0|Z)\operatorname{Pr}^*(Y^1=y^1|Z).\nonumber
\end{eqnarray}
\end{subequation}
Equation (\ref{eq:faca}) ``ties'' the induced distribution to the source
population distribution, thus allowing for samples from the former to
be used for inference on the latter. Equation (\ref{eq:facb}) determines the
correlation structure of the data, which is crucial for correct\vadjust{\goodbreak}
standard error computations. In twin studies, (\ref{eq:faca}) and
(\ref{eq:facb}) do not necessarily hold (see Section~\ref{sec:twins}),
but are assumed throughout the paper.

The induced marginal distribution of $Z$ is determined by the type of
matching. Under exposure-driven matching, the induced marginal
distribution of $Z$ equals the source population distribution of $Z$
among the exposed, that is,
$\operatorname{Pr}^*(Z)=\operatorname{Pr}(Z|X=1)$. In twin studies
restricted to the exposure--discordant pairs, we have that
$\operatorname{Pr}^*(Z)=\operatorname{Pr}(Z|\mbox{discordant in }X)$.

When $Z$ is observed (as in regular matched studies), the likelihood contribution for pair $i$ is
\begin{eqnarray*}
&&\operatorname{Pr}^*(Y^0_i=y^0_i,Y_i^1=y^1_i,Z_i)
\\
&&\quad=\prod_{x=0}^1\operatorname{Pr}(Y=y^x_i|X=x,Z_i)\operatorname{Pr}^*(Z_i),
\end{eqnarray*}
so that the likelihood for the whole data set becomes equal to
\[
\prod_{i=1}^n\prod_{x=0}^1\operatorname{Pr}(Y=y^x_i|X=x,Z_i)\operatorname{Pr}^*(Z_i).
\]
When $Z$ is unobserved (as in twin studies), the likelihood contribution for pair $i$ is
\begin{eqnarray*}
&&E^*_{Z_i}\{\operatorname{Pr}^*(Y^0_i=y^0_i,Y_i^1=y^1_i|Z_i)\}
\\
&&\quad=E^*_{Z_i}
\Biggl\{\prod_{x=0}^1\operatorname{Pr}(Y=y^x_i|X=x,Z_i) \Biggr\},
\end{eqnarray*}
so that the the likelihood for the whole data set becomes equal to
\[
\prod_{i=1}^nE^*_{Z_i}
\Biggl\{\prod_{x=0}^1\operatorname{Pr}(Y=y^x_i|X=x,Z_i) \Biggr\}.
\]
We note that marginally (over $Z$), $Y^0$ and $Y^1$ are associated through the common value
of $Z$; the strong\-er~conditional association between $Y$ and $Z$, giv\-en~$X$,~the stronger
marginal association between $Y^0$\break and~$Y^1$.

\section{Analysis Methods}
\label{sec:analysis}

In this section we describe and compare the most common analysis
methods for matched cohorts. We emphasize that all these methods can in
principle be used to analyze the exposure--discordant pairs in twin
studies as well. However, the explicit regression model (Section~\ref{sec:explicit}) requires $Z$ to be observed, which is typically not
the case in twin studies.

\subsection{Regression Model Explicitly Involving $Z$}
\label{sec:explicit}

A straightforward way to adjust for $Z$ is to fit a~regression model for $Y$, given $X$ and $Z$, for example,
\begin{equation}
\label{eq:mod1}
\hspace*{14pt}\operatorname{logit}\{\operatorname{Pr}(Y=1|X,Z;\psi_c,\gamma)\}=b(Z;\gamma)+\psi_c
X,\hspace*{-14pt}
\end{equation}
where $b(Z;\gamma)$ is an explicitly specified parametric function of
$Z$, typically a linear function $\gamma^TZ$ for continuous $Z$. We
refer to a regression model for~$Y$, given $X$ and $Z$, as
``explicit.'' Under model (\ref{eq:mod1}),
$\log\{\mathit{OR}_c(Z)\}=\psi_c$, so that the condition in
(\ref{eq:const}) is met. This restriction is not crucial though; in
principle we can add arbitrary interaction terms between~$X$ and any of
the components of $Z$. Maximum likelihood estimates (MLEs) of
($\psi_c,\gamma$) are obtained by maximizing the conditional (given
$Z$) likelihood
\begin{eqnarray}
\label{eq:lik1}\quad
&&\prod_{i=1}^n\operatorname{Pr}^*(Y^0_i=y^0_i,Y_i^1=y^1_i|Z_i)\nonumber
\\[-8pt]\\[-8pt]
&&\quad=\prod_{i=1}^n\prod_{x=0}^1\operatorname{Pr}(Y=y^x_i|X=x,Z_i;\psi_c,\gamma),\nonumber
\end{eqnarray}
where the equality follows from (\ref{eq:faca}) and (\ref{eq:facb}). If
(\ref{eq:facb}) is violated, then $Y^1$ and $Y^0$ are not conditionally
independent, given $Z$, and the right-hand side of~(\ref{eq:lik1}) is
not a proper likelihood. However, if (\ref{eq:faca}) holds (and model
(\ref{eq:mod1}) is correct), then each separate term
$\operatorname{Pr}(Y=y^x_i|X=x,Z_i;\psi_c,\gamma)$ in (\ref{eq:lik1})
equals the true marginal (over $Y_i^{1-x}$) likelihood
$\operatorname{Pr}(Y_i^x=y_i^x|Z_i)$. It follows that the obtained
estimate of $\psi_c$ is consistent under (\ref{eq:faca}), regardless of
whether (\ref{eq:facb}) holds or not.

\subsubsection{Disadvantages}
\label{sec:rtrt}
\begin{longlist}
\item[(1)] If $Z$ is high dimensional, it may be difficult
  to well specify the function $b(Z;\gamma)$.
\item[(2)]  If $Z$ is not directly observed, as in twin studies,
explicit specification of $b(Z;\gamma)$ is not possible.
\item[(3)] In principle, explicit regression models can be adapted for risk differences and
risk ratios, by using identity links or the log links, respectively. However,
absolute risks and logarithms thereof are, unlike log odds, restricted to ranges
$(0,1)$ and $(0,\infty)$, respectively. Thus, models utilizing identity links or log
links have to be fitted under these restrictions, which can be rather inconvenient, or
 they may produce estimates which are outside the supported ranges.
\end{longlist}

\subsection{Conditional Logistic Regression}
\label{sec:cond} Conditional logistic regression mitigates the
problems with an explicit specification of $b(Z;\gamma)$. In
conditional logistic regression, the function $b(Z;\gamma)$ in~(\ref{eq:mod1})
is replaced with a scalar pair-specific parameter~$b$:
\begin{equation}
\label{eq:mod2}
\operatorname{logit}\{\operatorname{Pr}(Y=1|X,Z)\}=b+\psi_c X.
\end{equation}
Nothing is assumed about $b$, and thus the risk for model
misspecification in $b(Z;\gamma)$ is avoided. A MLE of $\psi_c$ is obtained
by conditioning on $Y^0_i+Y^1_i$, for each pair $i$, and maximizing
the resulting conditional likelihood, which under (\ref{eq:faca})
and (\ref{eq:facb}) is given by
\begin{equation}
\label{eq:condlik}
\prod_{i:y_i^0\neq y_i^1}\frac{e^{\psi_c y^1_i}}{1+e^{\psi_c }}.
\end{equation}
Since the conditional likelihood (\ref{eq:condlik}) does not involve~$b$
(or $Z$), it can be used, even if $Z$ is not directly observed,
as in twin studies. The MLE of $\psi_c$ obtained by maximizing (\ref{eq:condlik}) is given by
\begin{equation}
\label{eq:condest} \hat{\psi}_{\mathit{c.clr}}=\log(U/V),
\end{equation}
 with standard error $s.e.\{\hat{\psi}_{\mathit{c.clr}}\}=\sqrt{U^{-1}+V^{-1}}$.

\subsubsection{Disadvantages}
\begin{longlist}
\item[(1)] The constant odds ratio assumption (\ref{eq:const}) is crucial in conditional logistic regression.
If an interaction term is included between $b$ and $X$ in model~(\ref{eq:mod2}), then $b$ cannot be eliminated
by conditioning arguments. If (\ref{eq:const})
 is violated, then $\hat{\psi}_{\mathit{c.clr}}$ converges to a~weighted average of the $Z$-specific odds ratios;
 see Section~\ref{sec:std}.
\item[(2)] $\hat{\psi}_{\mathit{c.clr}}$ is generally inconsistent if (\ref{eq:facb})
is violated. There is an important exception. Define the null hypothesis
\begin{equation}
\label{eq:h0} \mathrm{H}_{0}\dvtx \quad\mbox{(\ref{eq:const}) holds, with
}\psi_c=0.
\end{equation}
In Appendix \ref{sec:app4} we show that $\hat{\psi}_{\mathit{c.clr}}$ converges
to 0 under $\mathrm{H}_{0}$ and (\ref{eq:faca}), regardless of whether
(\ref{eq:facb}) holds or not. \item[(3)] Conditional logistic
regression cannot be used for other measures of association than the
log odds ratio, since for other links than the logit link, $b$ cannot
be eliminated by conditioning arguments.
\end{longlist}

\subsection{Mixed Model}\label{sec:mixed}

In the mixed model approach, $b$ is assumed to be random, with a
specified parametric distribution $\operatorname{Pr}^*(b;\theta)$. The
MLE of $(\psi_c,\theta)$ is obtained  by maximizing the marginal (over
$b$) likelihood
\begin{eqnarray}
\label{eq:mixed}
&&\prod_{i=1}^nE^*_{Z_i}\{\operatorname{Pr}^*(Y_i^0=y^0_i,Y_i^1=y^1_i|Z_i)\}\hspace*{-15pt}\nonumber\\[-2pt]\\[-14pt]
&&\quad= \prod_{i=1}^nE^*_{b_i}  \Biggl[   \Biggl\{
\prod_{x=0}^1\operatorname{Pr}(Y=y^x_i|X=x,b_i;\psi_c)  \Biggr\};
\theta \Biggr],\nonumber\hspace*{-15pt}
\end{eqnarray}
where the equality follows from (\ref{eq:faca}) and (\ref{eq:facb}),
and the expectation on the right-hand side is taken over
$\operatorname{Pr}^*(b;\theta)$. \citet{Neuhaus2} showed that the mixed
model estimate of $\psi_c$ is identical to $\hat{\psi}_{\mathit{c.clr}}$, under
mild conditions. This implies that the two methods are equally
efficient, and that the mixed model is robust against misspecification
of $\operatorname{Pr}^*(b;\theta)$.

\subsubsection{Disadvantages}
\begin{longlist}
\item[(1)] The constant odds ratio assumption (\ref{eq:const}) is crucial in the mixed model.
\citet{Neuhaus2} showed that the mixed model is saturated, under mild conditions, so that an
interaction term be-\break tween~$b$~and $X$ would lead to identifiability~\mbox{problems}.
\item[(2)] The mixed model estimate of $\psi_c$ is generally inconsistent if (\ref{eq:facb}) is violated.
\item[(3)] In principle, the mixed model can be adapted for risk differences and risk ratios,
by using identity links or the log links, respectively. In practice, these adaptations require
that the model is fitted under restrictions, or it may produce estimates outside the supported ranges.
\item[(4)] Explicit maximization of the likelihood in (\ref{eq:mixed}) requires numerical
techniques. This makes the meth\-od less transparent and relatively computer-intensive.
\end{longlist}

\subsection{Exposure--Discordant Crude Analysis}\label{sec:std}

The methods described in Sections \ref{sec:explicit}--\ref{sec:mixed}
all target the conditional odds ratio, $\mathit{OR}_c(Z)$. Matched~da\-ta
can also be used to estimate a standardized odds ratio. Let $n_{yx}$
denote the number of subjects in the sample with $Y=y$ and $X=x$, so
that $n_{00}=U+W$, $n_{01}=V+W$, $n_{10}=V+T$ and $n_{11}=U+T$. Under
(\ref{eq:faca}) we have that
$\operatorname{Pr}^*(Y^x=y^x)=E^*_Z\{\operatorname{Pr}(Y=y^x|X=x,Z)\}$,
that is, $\operatorname{Pr}^*(Y^x=y^x)$ equals the probability of
$Y=y^x$ given $X=x$, standardized to $\operatorname{Pr}^*(Z)$. Thus,
under (\ref{eq:faca}) a consistent estimate of $\psi_s$ is given by the
crude log odds ratio
\begin{equation}
\label{eq:psiest} \hat{\psi}_{\mathit{s.crude}}=\log  \biggl(
\frac{n_{11}n_{00}}{n_{01}n_{10}}  \biggr).
\end{equation}

The standard error of $\hat{\psi}_{\mathit{s.crude}}$ (see Appendix \ref{sec:app2}) is given by
\begin{equation}
\label{eq:se}
\hspace*{14pt}\sqrt{n_{11}^{-1}+n_{01}^{-1}+n_{10}^{-1}+n_{00}^{-1}-2n\frac{nT-n_{11}n_{10}}{n_{11}n_{00}n_{01}n_{10}}}.\hspace*{-14pt}
\end{equation}
The first four terms under the square root sign can be recognized from the usual standard error formula
for a log odds ratio, and the fifth term is an adjustment for non-i.i.d. observations.

We remind the reader that the interpretation of~$\psi_s$ depends on
what distribution\vadjust{\goodbreak} of $Z$ that $\psi_s$ is standardized to. Under
exposure-driven matching,\break
$\operatorname{Pr}^*(Z)=\operatorname{Pr}(Z|X=1)$ so that $\psi_s$ is
standardized to the distribution of $Z$ among the exposed. In a~twin
study, $\operatorname{Pr}^*(Z)=\operatorname{Pr}(Z|\mbox{discordant in
}X)$ so that $\psi_s$ is standardized to the distribution of $Z$ among
the exposure--discordant pairs.

\subsubsection{Advantages}

One potential disadvantage of the exposure--discordant crude analysis
is that it estimates a parameter that is rather nonstandard. In the
simple scenario that we consider (i.e., 1:1 matching and no additional
covariate adjustments) the exposure--discordant crude analysis does not
suffer from any of the other disadvantages listed in Sections
\ref{sec:explicit}--\ref{sec:mixed}. The relative advantages of the
expo\-sure--discordant crude analysis are threefold:
\begin{longlist}
\item[(1)] The exposure--discordant crude analysis relies on fewer
assumptions than the other methods. Specifically, it does not rely on
assumptions (\ref{eq:const})\break and~(\ref{eq:facb}).
\item[(2)] The exposure--discordant crude analysis is\break computationally simple.
\item[(3)] In the exposure--discordant crude analysis, the standardized
probabilities $\operatorname{Pr}^*(Y^0=1)$ and\break
$\operatorname{Pr}^*(Y^1=1)$ can be estimated separately, and can
subsequently be used to construct any standardized measure of the
$X$--$Y$ association, for example, risk difference or risk ratio. For
this reason, the exposure--discordant crude analysis easily extends to
nonbina\-ry outcomes as well. For survival outcomes, for instance, an
exposure--discordant crude analysis can~be used to produce standardized
Kaplan--Meier curves.
\end{longlist}

\subsubsection{A closer comparison with conditional logistic regression}
Because $\psi_s$ and $\psi_c$ are different parameters, it is not
meaningful to compare the methods in  Sections
\ref{sec:explicit}--\ref{sec:mixed} with the exposure--discordant crude
analysis in terms of efficiency of estimates. However, we can make a
meaningful comparison in terms of statistical power. Define the null
hypothesis
\begin{equation}
\label{eq:hostar} \mathrm{H}_{0}^*\dvtx  \quad\psi_s=0.
\end{equation}
It is easy to show that $\mathrm{H}_{0}$ in (\ref{eq:h0}) implies
$\mathrm{H}_{0}^*$, regardless of whether (\ref{eq:faca}) and
(\ref{eq:facb}) hold or not. If both (\ref{eq:faca}) and
(\ref{eq:facb}) hold, then a Wald test of $\mathrm{H}_{0}$ is based on
the statistic $T_c=\hat{\psi}_{\mathit{c.clr}}/s.e.(\hat{\psi}_{\mathit{c.clr}})$. If
(\ref{eq:faca}) holds, then a Wald test of $\mathrm{H}_{0}^*$ is based
on the statistic $T_s=\hat{\psi}_{\mathit{s.crude}}/s.e.(\hat{\psi}_{\mathit{s.crude}})$.
In Appendix \ref{sec:app4} we show that $T_c$ and $T_s$ are
asymptotically equal. It immediately follows that the two Wald tests
have the same asymptotic power, for any fixed alternative.

One potential argument against the exposure--\break discordant crude analysis
is that it does not inform us about the exposure effect in the source
population. Under exposure-driven matching (and no confounders apart
from $Z$), $\psi_s$ is a causal effect in a fictitious population where
$Z$ is distributed as among the exposed. In a twin study restricted to
the expo\-sure--discordant pairs (and no confounders apart\break from~$Z$),
$\psi_s$ is a causal effect in a fictitious population where~$Z$ is
distributed as among the exposure--discordant pairs. The effect in
these fictitious populations may differ from the effect in the source
population, and it is not always obvious whether these fictitious
population effects are relevant targets for inference. However, a
closer examination shows that a similar argument can be used against
the methods that target $\psi_c$ as well, and in particular against
conditional logistic regression. Conditional logistic regression relies
on the constant odds ratio assumption~(\ref{eq:const}). This is a~very
strong assumption, which in any real scenario is most likely violated,
to some extent. Regardless of whether (\ref{eq:const}) holds  or not,
$\hat{\psi}_{\mathit{c.clr}}$ converges to\vspace*{-8pt}

{\fontsize{9.5}{11.5}{\selectfont{
\begin{eqnarray}
\label{eq:or}
&&\hspace*{-5pt}\log  \biggl\{ \frac{\operatorname{Pr}^*(Y^1=1,Y^0=0)}{\operatorname{Pr}^*(Y^0=1,Y^1=0)}  \biggr\}\nonumber\\
&&\hspace*{-5pt}\quad =\log  \biggl[\frac{E^*_Z\{\operatorname{Pr}^*(Y^1=1,Y^0=0|Z)\}}{E^*_Z\{\operatorname{Pr}^*(Y^0=1,Y^1=0|Z)\}}
\biggr]\nonumber\\
&&\hspace*{-5pt}\quad \stackrel{\mathrm{(\ref{eq:faca}),(\ref{eq:facb})}}{=}\log  \biggl[ \frac{E^*_Z\{\operatorname{Pr}(Y=1|X=1,Z)\operatorname{Pr}(Y=0|X=0,Z)\}}{E^*_Z\{\operatorname{Pr}(Y=1|X=0,Z)\operatorname{Pr}(Y=0|X=1,Z)\}}  \biggr]\nonumber\\
&&\hspace*{-5pt}\quad =\log[E^*_Z\{W(Z)\mathit{OR}_c(Z)\}],\nonumber\\
\end{eqnarray}}}}%
where
\begin{eqnarray*}
&&W(Z)
\\
&&\quad=\frac{\operatorname{Pr}(Y=1|X=0,Z)\operatorname{Pr}(Y=0|X=1,Z)}{E^*_Z\{\operatorname{Pr}(Y=1|X=0,Z)\operatorname{Pr}(Y=0|X=1,Z)\}}.
\end{eqnarray*}
In (\ref{eq:or}), the average is taken over $\operatorname{Pr}^*(Z)$,
that is, the same distribution of $Z$ as being standardized to in the
exposure--discordant crude analysis.  Thus, if (\ref{eq:const}) is
violated, then conditional logistic regression does not inform the
analyst about exposure effects outside the fictitious population
characterized by $\operatorname{Pr}^*(Z)$, to any wider extent than the
exposure--discordant crude analysis. Furthermore, whereas $\psi_s$ has
a clear interpretation as a population causal effect (when there are no
confounders except $Z$), the weighted average in (\ref{eq:or}) does not
have any such simple interpretation.

An analyst is always at the liberty to assume a priori that
(\ref{eq:const}) holds. But equally well, the analyst may assume that
the effect in the fictitious population, characterized by
$\operatorname{Pr}^*(Z)$, is equal to the effect in the source
population, characterized by $\operatorname{Pr}(Z)$. Neither of these
assumptions is stronger than the other, since neither of them implies
the other. Furthermore, with paired data and $Z$ being unobserved (as
in twin studies), these assumptions are both untestable.

Although our focus is on cohort studies, we end this section by making
a comparison with case control studies. A matched case control study is
designed analogously to a matched cohort study, but the roles of
exposure and outcome are ``switched'' in the sampling scheme; see
Section~\ref{sec:introduction}. Thus, in a~match\-ed case control study
the crude sample log odds ratio consistently estimates the standardized
log odds ratio\vspace*{-8pt}

{\fontsize{10.4}{12.4}{\selectfont{
\begin{eqnarray}
\label{eq:orstt}  &&\hspace*{-4pt}\log  \biggl[
\frac{E^*_Z\{\operatorname{Pr}(X=1|Y=1,Z)\}E^*_Z\{\operatorname{Pr}(X=0|Y=0,Z)\}}{E^*_Z\{\operatorname{Pr}(X=0|Y=1,Z)\}E^*_Z\{\operatorname{Pr}(X=1|Y=0,Z)\}}
\biggr] ,\nonumber\\&&
\end{eqnarray}}}}%
where $\operatorname{Pr}^*(Z)=\operatorname{Pr}(Z|Y=1)$. In contrast to
conditional odds ratios, standardized odds ratios are not symmetrical.
That is, the log odds ratio in~(\ref{eq:orstt}), in which $X$ appears
to the left of the conditioning sign, cannot be written as $\psi_s$, in
which $X$ appears to the right of the conditioning sign. Hence, the log
odds ratio in~(\ref{eq:orstt}) has no simple interpretation as a causal
effect of $X$ on $Y$ on the log odds ratio scale, even if there are no
confounders apart from $Z$.

\section{Analysis of twin data}
\label{sec:twins}

In contrast to a regular 1:1 matched cohort study, a twin cohort also
contains pairs that are concordant in the exposure. In this section we
describe three common methods to incorporate the exposure--con\-cordant
pairs in the analysis.

To deal with twin studies we extend the notation slightly.  Let
$X_{ij}$ and $Y_{ij}$ denote $X$ and $Y$ for twin $j$ in pair $i$,
$j\in (1,2)$. We suppress the index $i$ when not needed, so that $X_j$
and $Y_j$ denote $X$ and $Y$ for twin $j$, $j\in (1,2)$, within an
arbitrary pair $i$. As before, $Z_i$ represents all the unobserved
factors that are common within a twin pair. As discussed in Section~\ref{sec:introduction}, the exposure--discordant pairs in a twin cohort
can be viewed as a 1:1 matched cohort. However, some care must be
taken. All methods discussed in Section~\ref{sec:analysis} rely on
assumption (\ref{eq:faca}), and conditional logistic regression
(Section~\ref{sec:cond}) and mixed models\vadjust{\goodbreak} (Section~\ref{sec:mixed})
rely in addition on assumption~(\ref{eq:facb}). For an
exposure--discordant twin pair we have that
\begin{eqnarray}
\label{eq:factwins}
&&\hspace*{12pt}\operatorname{Pr}^*(Y^0=y^0,Y^1=y^1|Z)\nonumber\hspace*{-12pt}
\\[-8pt]\\[-8pt]
&&\hspace*{12pt}\quad=\operatorname{Pr}(Y_j=y^0,Y_{j'}=y^1|X_j=0,X_{j'}=1,Z).\nonumber\hspace*{-12pt}
\end{eqnarray}
The right-hand side of (\ref{eq:factwins}) can be factorized into
$\operatorname{Pr}(Y_j=y^0|X_j=0,Z)\operatorname{Pr}(Y_{j'}=y^1|X_{j'}=1,Z)$
if
\begin{subequation}
\begin{equation}
\label{eq:factwins1}  Y_j\perp X_{j'}|(X_j,Z)\end{equation}
 and
 \begin{equation}
 \label{eq:factwins2}
Y_1\perp Y_2|(X_1,X_2,Z).
\end{equation}
\end{subequation}
Thus, the analogs to (\ref{eq:faca}) and (\ref{eq:facb}) for twin data
are given by (\ref{eq:factwins1}) and (\ref{eq:factwins2}),
respectively. Under (\ref{eq:factwins1}), (\ref{eq:faca}) holds, so
that the explicit model (Section~\ref{sec:explicit}) and the
exposure--discordant crude analysis (\ref{sec:std}) are valid when
applied to the exposure--discordant pairs. We note though that it is
typically not possible to fit an explicit model to twin data, since $Z$
is typically unobserved. If, in addition,  (\ref{eq:factwins2}) holds,
then~(\ref{eq:facb}) holds as well, and all methods in Section~\ref{sec:analysis}
are valid when applied to the exposure--discordant pairs.\looseness=1

Potentially, (\ref{eq:factwins1}) could be violated if $X_{j'}$
has\break
a~caus\-al effect on $Y_j$, that is, if the exposure for one twin affects
the outcome for the other twin.  Similarly, (\ref{eq:factwins2}) could
be violated if $Y_{j'}$ has a causal effect on $Y_j$, that is, if the
outcome of one twin affects the outcome for the other twin.

\subsection{All-Pair Crude Analysis}
\label{sec:crude2}

Let $r_{yx}$ denote the number of subjects in the full (i.e., both
exposure--concordant and exposure--dis\-cordant pairs) sample with $Y=y$
and $X=x$. One simple way to make use of all twin pairs in the analysis
is to compute the crude sample log odds ratio
\begin{equation}
\label{eq:ty} \hat{\psi}_{m.crude}=\log  \biggl(
\frac{r_{11}r_{00}}{r_{01}r_{10}}  \biggr),
\end{equation}
which consistently estimates the marginal log odds ratio $\psi_m$.
Thus, unlike the exposure--discordant crude analysis (Section~\ref{sec:std}), the all-pair crude analysis does not adjust for
confounding by $Z$. The standard error of $\hat{\psi}_{m.crude}$ is
rather complicated, due to the paired nature of the data. In Appendix
\ref{sec:app2} we provide an analytic expression for the standard
error. We note that the standard error can also be computed
numerically, through Generalized Estimating Equation (GEE) procedures,
which are implemented in most common statistical softwares.

\subsection{Decomposition into Within- and Between-Effects}
\label{sec:withinbetween}

In twin studies with continuous exposures and outcomes, a popular regression model is
\begin{eqnarray}
\label{eq:WB}
\quad E(Y_j|X_j,X_{j'})&=&\beta_0+\beta_{\mathrm{W}}(X_j-\bar{X})+\beta_{\mathrm{B}}\bar{X}\nonumber\\[-8pt]\\[-8pt]
&=&\beta_0+\beta_{\mathrm{W}}X_j+\beta_{\mathrm{B}}'\bar{X},\nonumber
\end{eqnarray}
with $\bar{X}=\frac{X_1+X_2}{2}$ and
$\beta_{\mathrm{B}}'=\beta_{\mathrm{B}}-\beta_{\mathrm{W}}$
\citep{Carlin}. In (\ref{eq:WB}), the pair-specific mean $\bar{X}$ is
thought of as conveying information about the confounders~$Z$, which
are not observed, but constant within each pair. Thus, the parameter
$\beta_{\mathrm{B}}$ is thought of as quantifying the strength of
confounding,  a ``between effect,'' and the parameter
$\beta_{\mathrm{W}}$ is thought of as quantifying the adjusted $X$--$Y$
association, a ``within effect.'' When $X$ and $Y$ are binary, a
natural analog to (\ref{eq:WB}) is
\begin{eqnarray}
\label{eq:WB2}
&&\operatorname{logit}\{\operatorname{Pr}(Y_j=1|X_j,X_{j'})\}\nonumber
\\[-8pt]\\[-8pt]
&&\quad=\beta_0+\beta_{\mathrm{W}}X_j+\beta_{\mathrm{B}}'\bar{X}.\nonumber
\end{eqnarray}
To see the connection with the methods described in this paper, note that
\begin{eqnarray*}
\label{eq:betapsi}
\beta_{\mathrm{W}}&=&\operatorname{logit}\{\operatorname{Pr}(Y_j=1|X_j=1,X_{j'}=0)\}\\
&&{}-\operatorname{logit}\{\operatorname{Pr}(Y_j=1|X_j=0,X_{j'}=1)\}\\
&=&\operatorname{logit}[E\{\operatorname{Pr}(Y_j=1|X_j=1,X_{j'}=0,Z)|
\\
&&\hspace*{99pt}{}X_j=1,X_{j'}=0\}]\\
&&{}-\operatorname{logit}[E\{\operatorname{Pr}(Y_j=1|X_j=0,X_{j'}=1,Z)|\\
&&\hspace*{111pt}{}X_j=0,X_{j'}=1\}]\\
&=&\operatorname{logit}[E^*\{\operatorname{Pr}(Y_j=1|X_j=1,Z)\}]\\
&&{}-\operatorname{logit}[E^*\{\operatorname{Pr}(Y_j=1|X_j=0,Z)\}]\\
&=&\psi_s,
\end{eqnarray*}
where $\operatorname{Pr}^*(Z)=\operatorname{Pr}(Z|X_1\neq X_2)$, and
the third equality follows from assumption (\ref{eq:factwins1}). Thus,
the within-effect $\beta_W$ is identical to the log odds ratio
standardized to the distribution of $Z$ among the exposure--discordant
pairs. This argument shows that the decomposition into within- and
between-effects is a legitimate method for binary exposures, which was
questioned by \citet{Carlin}.

When $X$ is binary, $\bar{X}$ can only take values 0, 0.5 and~1. Thus,
it is feasible to replace the linear term
$\beta_0+\beta_{\mathrm{B}}'\bar{X}$ in (\ref{eq:WB2}) with one
parameter for each level of $\bar{X}$, that is,
\begin{equation}
\label{eq:WB3}
\hspace*{14pt}\operatorname{logit}\{\operatorname{Pr}(Y_j=1|X_j,X_{j'})\}=\beta_{\mathrm{W}}X_j+m(\bar{X}),\hspace*{-14pt}
\end{equation}
with
\begin{eqnarray}
\label{eq:m}
m(\bar{X})&=&\beta_0\mathbf{1}(\bar{X}=0)\nonumber\\[-8pt]\\[-8pt]
&&{}+\beta_{0.5}\mathbf{1}(\bar{X}=0.5)+\beta_1\mathbf{1}(\bar{X}=1).\nonumber
\end{eqnarray}
It is easy to show that the model in (\ref{eq:WB3}) is saturated (i.e.,
imposes no restrictions on $\operatorname{Pr}(Y_j|X_1,X_2)$, which
implies that the MLE of $\beta_{\mathrm{W}}$ based on (\ref{eq:WB3}) is
identical to the crude sample log odds ratio in (\ref{eq:psiest}).

\subsection{Mixed Model}
\label{sec:mixedall}

The model in (\ref{eq:mod2}) can be fitted to all pairs, assuming a
parametric distribution of $b$ indexed with~$\theta$. Parameter
estimates are obtained by maximizing the marginal (over $b$) likelihood
\begin{eqnarray}
\label{eq:mixedfull}
&&\hspace*{-3pt}\prod_{i=1}^nE^*_{Z_i|X_{i1},X_{i2}}\{\operatorname{Pr}(Y_{i1}=y_{i1},Y_{i2}=y_{i2}|X_{i1},X_{i2},Z_i)|\nonumber
\\
&&\hspace*{189pt}{}X_{i1},X_{i2}\}\nonumber\\[-8pt]\\[-8pt]
&&\hspace*{-3pt}\quad =\prod_{i=1}^nE^*_{b_i|X_{i1},X_{i2}}  \Biggl[   \Biggl\{
\prod_{j=1}^2\operatorname{Pr}(Y_{ij}=y_{ij}|X_{ij},b_i;\psi_c)
\Biggr\}\Big|\nonumber\\
&&\hspace*{167pt}\quad{}  X_{i1},X_{i2};\theta  \Biggr].\nonumber
\end{eqnarray}
This approach, however, is associated with a severe problem which is
often overlooked. Typically, the distribution of $b$ is specified to
not depend on $(X_1,X_2)$, for example, a normal distribution with
fixed but unspecified mean and variance. However, from the expression
in (\ref{eq:mixedfull}) it is clear that this procedure only produces a
proper likelihood under the additional assumption that $b\perp
(X_1,X_2)$. In standard textbooks, this assumption is often stated
without justification or interpretation (e.g., Fitzmaurice et al., \citeyear{Fitzmaurice},
page~329). Since $b$ is supposed to represent the potential confounders
$Z$, we would not generally expect that $b\perp (X_1,X_2)$. Indeed, if
$Z$ (and thus~$b$) is independent of $(X_1,X_2)$, it cannot be
a~confounder, and there is no need to adjust for $Z$ in the first place.
We note that in matched cohort studies, $(X_1,X_2)$ is constant and
equal to $(0,1)$ for all pairs, so that an association between $b$ and
$(X_1,X_2)$ is ruled out by design. When $b$ is associated with
$(X_1,X_2)$, the aforementioned procedure can yield severely biased
estimates (\citecs{Neuhaus}; \citecs{Neuhaus3}). In general, the proper
marginal likelihood is obtained by averaging over a specified
distribution $\operatorname{Pr}(b|X_1,X_2)$ for each pair. This
procedure can be very computer intensive, and cannot be carried out
with standard software. As noted by \citet{Neuhaus} and
\citet{Neuhaus3}, there is a simple solution to this problem. Suppose
that given $(X_1,X_2)$, $b$ has a normal distribution where the mean,
but not the variance, depends on $(X_1,X_2)$. Without loss of
generality, we can formulate this as
\begin{equation}
\label{eq:norm}
b=d+m(\bar{X}),
\end{equation}
where $m(\bar{X})$ is defined in (\ref{eq:m}) and $d|X_1,X_2\sim\break
N(0,\sigma^2)$.  Under (\ref{eq:norm}), model (\ref{eq:mod2})
translates to
\begin{equation}
\label{eq:mod3} \hspace*{14pt}\operatorname{logit}\{
\operatorname{Pr}(Y_j=1|X_j,Z)\}=d+\psi_c
X_j+m(\bar{X}),\hspace*{-14pt}
\end{equation}
where $d\perp (X_1,X_2)$ by construction. The model\break in~(\ref{eq:mod3})
can be fitted with standard mixed model software. By comparing the
model in~(\ref{eq:mod3}) with the model in~(\ref{eq:WB3}), we see that
the solution proposed by \citet{Neuhaus} and \citet{Neuhaus3} can be
thought of as combining a mixed model with a within-between
decomposition.

\begin{table*}
\caption{Simulation results for $\psi_c=0$, $\phi=4$}
\label{tab:tab2}
\begin{tabular}{@{}lcccc@{}}
\hline
\textbf{Analysis method} & \textbf{Target parameter} & \textbf{Mean est} & \textbf{Emp s.e.}& \textbf{Th s.e.} \\
\hline
1. Explicit & $\psi_c=0$ & 0.00 & 0.13 & 0.13 \\
2. Cond log reg & $\psi_c=0$ & 0.00 & 0.13 & 0.13 \\
3. Mixed discordant &  $\psi_c=0$ & 0.00 & 0.13 & 0.13 \\
4. Crude discordant & $\psi_s=0$ &  0.00 & 0.11 & 0.11 \\
5. Crude all & $\psi_m=1.28$ & 1.28 & 0.08 & 0.08 \\
6. Mixed all & $\psi_c=0$ &  0.00 & 0.12 & 0.12 \\
\hline
\end{tabular}
\end{table*}

\begin{figure*}[b]
\includegraphics{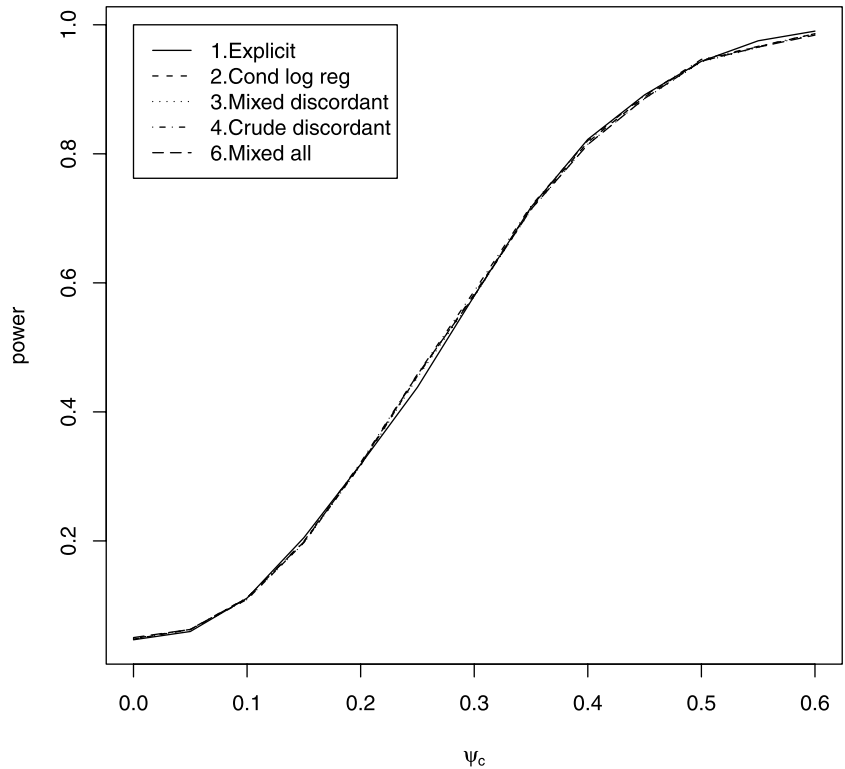}

\caption{Simulation results for $\psi_c\in (0,0.6)$,
$\phi=4$.} \label{fig:powerpsi}
\end{figure*}

\citet{Neuhaus} and \citet{Neuhaus3} observed that for various
scenarios, the estimate of $\psi_c$ obtained by combining a~mixed model
with a within-between decomposition is nearly identical to
$\hat{\psi}_{\mathit{c.clr}}$. \citet{Neuhaus3} gave a theoretical motivation
for this observation. We note that there are situations when the two
estimates may differ; see \citet{Brumback} for an example.

\section{Simulations}
\label{sec:sim}

\subsection{Part I: Efficiency and Power}
\label{sec:eff}

In this section we compare the performance of the methods described in
Sections \ref{sec:analysis} and \ref{sec:twins}, in terms of efficiency
and power. To enable a fair comparison, we analyze the simulated data
so that all assumptions hold, for each method respectively. In these
simulations, twin pairs were generated. We emphasize that this
simulation scheme covers matched data as well, since the
exposure--discordant twin pairs can be viewed as a matched cohort. For
each twin pair, the random variables ($X_1,X_2,b,Y_1,Y_2$) were
generated from the model
\fontsize{10pt}{\baselineskip}\selectfont
\makeatletter
\def\tagform@#1{\normalsize\maketag@@@{(\ignorespaces#1\unskip\@@italiccorr)}}
\makeatother
\begin{equation}
\label{eq:sim}
  \hspace*{12pt}\cases{
\displaystyle\frac{\operatorname{Pr}(X_1=1|X_2=0)}{\operatorname{Pr}(X_1=0|X_2=0)}=\frac{\operatorname{Pr}(X_2=1|X_1=0)}{\operatorname{Pr}(X_2=0|X_1=0)}\cr
\displaystyle\quad=\rho=\frac{1}{2},\cr
\displaystyle\frac{\operatorname{Pr}(X_1=1,X_2=1)\operatorname{Pr}(X_1=0,X_2=0)}{\operatorname{Pr}(X_1=1,X_2=0)\operatorname{Pr}(X_1=1,X_2=0)}=\phi,\cr
\displaystyle b|X_1,X_2  \sim  N  \{\theta\bar{X},1  \},\cr
\displaystyle Y_1\perp Y_2 |(X_1,X_2,b),\cr
Y_j\perp X_{j'} | (X_j,b),\cr
\displaystyle \operatorname{logit}\{\operatorname{Pr}(Y_j|X_j,b)\}=b+\psi_cX_j. }\hspace*{-12pt}
\end{equation}
\normalsize
We highlight a few aspects of the model in (\ref{eq:sim}):
\begin{longlist}
\item[(1)] Under model (\ref{eq:sim}), assumptions (\ref{eq:const}),
(\ref{eq:factwins1}), (\ref{eq:factwins2}) and (\ref{eq:norm}) all
hold.
\item[(2)] The restriction
$\operatorname{Pr}(X_1=1|X_2=0)=\operatorname{Pr}(X_2=1|X_1=0)$ in the
first row of (\ref{eq:sim}) follows by symmetry.
\item[(3)]  It may
appear natural to first specify a mar\-ginal distribution of $b$, then
specify a conditional distribution of ($X_1,X_2$), given $b$. The
reason for doing it the other way around is twofold. First, it allows
us to directly control the rate of exposure-discordance through~$\phi$.
Second, it allows us to easily formulate the distribution of $b$ given
($X_1,X_2$) in such a way that (\ref{eq:norm}) holds.
\item[(4)] It
follows from results in \citet{Chen} that the joint distribution of
$(X_1,X_2)$ is completely defined by $\rho$ and $\phi$. It also follows
that $\rho$ and $\phi$ are variation independent (i.e., the value of
$\rho$ does not restrict the value of $\phi$, and vice versa).
\item[(5)] The values of $\phi$ and $\theta$ determine the degree of
conditional association of $X_1$ and $X_2$, given $b$. It can be shown
(see Appendix \ref{sec:app1}) that for $\theta=2\sqrt{\log(\phi)}$,
$X_1\perp X_2|b$. For convenience, we have used
$\theta=2\sqrt{\log(\phi)}$ throughout. We note though that none of the
methods presented relies on this restriction.
\end{longlist}
In the first set of simulations, we used $\phi=4$ and $\psi_c=0$, that
is, the data were generated under $\mathrm{H}_{0}$ in~(\ref{eq:h0}).
For these values, $\psi_s=0$ and $\psi_m=1.28$, which implies a severe
degree of confounding. Further,\break $\operatorname{Pr}(X_1\neq X_2)=0.33$,
and $\operatorname{Pr}(X_1\neq X_2,Y_1\neq Y_2)= 0.11$. We generated
5000 samples, each of size $n=2000$. Each sample was analyzed with 6
different methods:
\begin{longlist}
\item[(1)] Explicit regression model
$\operatorname{logit}\{\operatorname{Pr}(Y=1|\break X, b)\}=\gamma_0+\gamma_1b+\psi_cX$
(Section~\ref{sec:explicit}). We remind the reader that for twin data,
$b$ (or rather, $Z$) is typically unobserved, which rules out the use
of an explicit model. For a regular matched cohort, the explicit model
is a viable choice. Thus, the model was only fitted to the
exposure--discordant pairs.
\item[(2)] Conditional logistic regression
(Section~\ref{sec:cond}).
\item[(3)] Mixed model fitted to the
exposure--discordant pairs (Section~\ref{sec:mixed}). We used the model
$\operatorname{Pr}(Y=1|\break X, b)=b+\psi_cX$, with $b|X_1\neq X_2 \sim
N(\theta,\sigma^2)$.
\item[(4)] Exposure--discordant crude analysis
(Sec-\break tion~\ref{sec:std}).
\item[(5)] All pair crude analysis (Section~\ref{sec:crude2}).
\item[(6)] Mixed model fitted to all pairs (Section~\ref{sec:mixedall}). We used the model
$\operatorname{Pr}(Y=1|X,b)=b+\psi_cX$, with $b|X_1,X_2 \sim
N(\theta\bar{X},\sigma^2)$.
\end{longlist}
Table \ref{tab:tab2} displays the mean (over samples) point
estimate, the empirical standard error and the mean theoretical
standard error for each analysis, respectively.
We note that all methods yield virtually unbiased
estimates of their target parameters. For all methods the mean theoretical standard error is identical to the
empirical standard error, to the second decimal.

To compare the methods in terms of their power to reject
$\mathrm{H}_{0}$, we carried out a second set of simulations. We used
$\phi=4$ and varied $\psi_c$ over the range $(0,0.6)$. For each value
of $\psi_c$, we drew 5000 samples of 2000 pairs each. Each sample was
analyzed using methods 1, 2, 3, 4, 6. Figure \ref{fig:powerpsi} displays
the empirical rejection probability (i.e., the power) for a Wald test
at 5\% significance level, for each method as a~function of~$\psi_c$.
We observe that the all methods have almost identical power, for the simulated
scenarios.\looseness=1

\begin{figure*}

\includegraphics{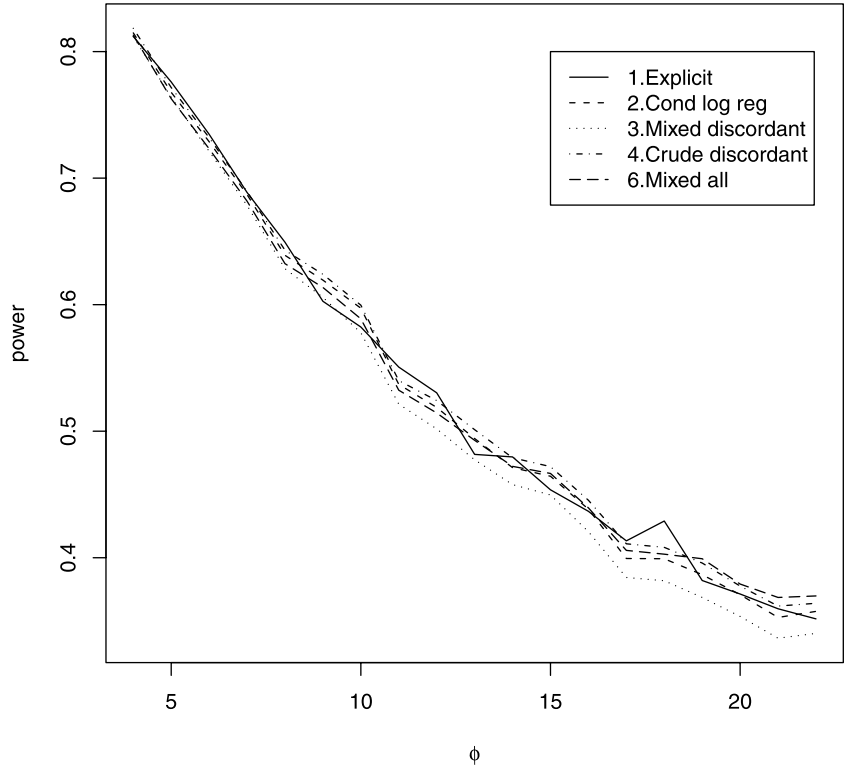}

\caption{Simulation results for $\psi_c=0.4$, $\phi\in (4,22)$.}
\label{fig:powerphi}
\end{figure*}

In a third set of simulations, we used $\psi_c=0.4$ and varied $\phi$
over the range $(4,22)$. These values correspond to the range
$(0.33,0.13)$ for $\operatorname{Pr}(X_1\neq X_2)$, and the range
$(0.11,0.03)$ for $\operatorname{Pr}(X_1\neq X_2,Y_1\neq Y_2)$. For
each value of $\phi$, we drew 5000 samples of 2000 pairs each. Each
sample was analyzed using methods 1, 2, 3, 4, 6. Figure \ref{fig:powerphi}
displays the power for each method as a function of $\phi$.
Again, we observe that there is almost no difference between the methods, in terms of power, even when the discordance rate is very low.

\begin{figure*}

\includegraphics{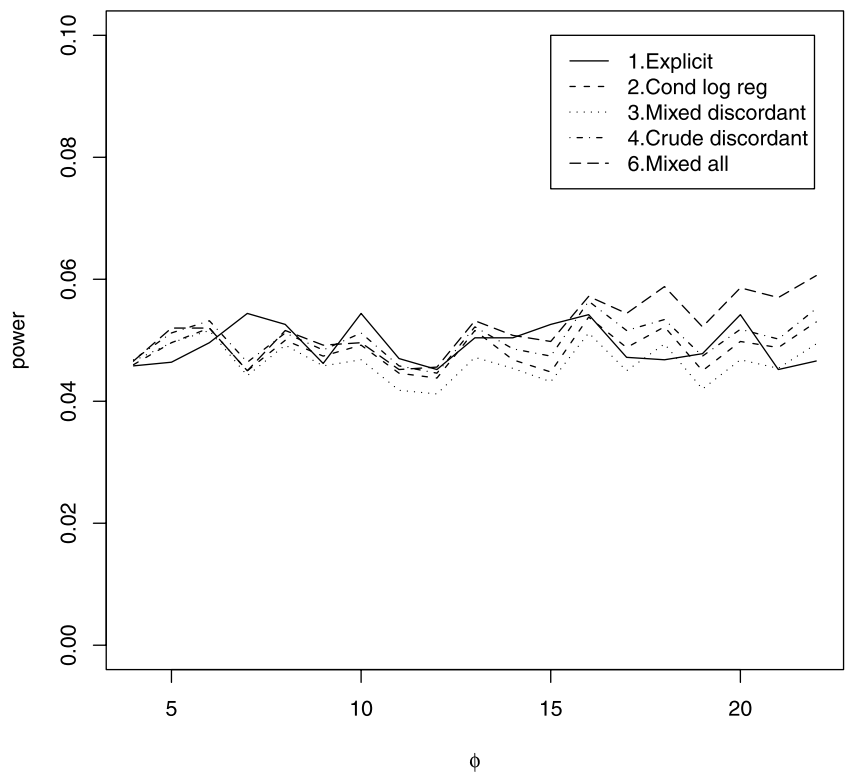}

\caption{Simulation
results for $\psi_c=0$, $\phi\in (4,22)$.} \label{fig:powerphi2}
\end{figure*}

Some care must be taken when interpreting power curves. In small
samples, parameter estimates can be biased, which may lead to an
increased probability of rejection, both under the alternative
hypothesis and under the null hypothesis. Thus, an increased power
under the alternative hypothesis may come at the cost of a violated
significance level under the null hypothesis. Figure \ref{fig:powerpsi}
shows that the nominal significance level ($=$~5\% at $\psi_c=0$) is
preserved for all methods when $\phi=4$. To confirm that the nominal
significance level is preserved across the range $\phi\in (4,22)$,
which generated the power curves in Figure \ref{fig:powerphi}, we
carried out a fourth set of simulations, using $\psi_c=0$ and varying
$\phi$ over the range $(4,22)$. For each value of $\phi$, we drew 5000
samples of 2000 pairs each. Each sample was analyzed using methods 1,
2, 3, 4, 6. Figure \ref{fig:powerphi2} displays the rejection
probability for each method as a function of $\phi$. We observe that
the rejection probability is close to 0.05, for all methods and all
values of $\phi$ in the simulated range.\looseness=1

Table \ref{tab:tab2} and Figure \ref{fig:powerpsi} indicate that
methods 1--4, and 6 are unbiased under the null hypothesis. Additional
simultions have confirmed that the methods are unbiased under various
alternative hypotheses as well (data not shown).

\subsection{Part II: Sensitivity to Underlying Assumptions}
\label{sec:rob}

In this section we demonstrate through examples that the explicit
model, conditional logistic regression and the mixed model, can yield
biased estimates, if their underlying assumptions are violated.

We first consider the assumption that $b\perp (X_1,X_2)$, which is often
made for mixed models; see Section~\ref{sec:mixedall}. Toward this end
we reanalyzed the 5000 simulated samples which generated Table
\ref{tab:tab2}, now fitting the mixed model
$\operatorname{Pr}(Y=1|X,b)=b+\psi_cX$ to all pairs, with $b|X_1,X_2
\sim N(\theta,\sigma^2)$. We obtained a mean estimate of $\psi_c$ equal
to 1.32, which is indeed biased as an estimate of the true value
$\psi_c=0$. We note that this mean estimate is very close to the
$\hat{\psi}_{m.crude}$ ($=1.28$) in Table \ref{tab:tab2}. This further
demonstrates that ignoring the association between $b$ and ($X_1,X_2$)
produces an estimate which is not adjusted for $Z$.

Next, we consider the independence assumption
(\ref{eq:facb})/(\ref{eq:factwins2}), which is a prerequisite for
conditional logistic regression and mixed models. Toward this end we
consider a simple scenario for which
\begin{equation}
\label{eq:ii}
(Y_1,Y_2)\perp Z|(X_1,X_2),
\end{equation}
so that $\psi_c=\psi_s=\psi_m$; see Section~\ref{sec:margcondstd}. We define
\begin{equation}
\label{eq:sim2}
  \hspace*{12pt}\cases{
\displaystyle\operatorname{Pr}(Y_j=1|Y_{j'}=0,X_j=1,X_{j'}=0)=p,\cr
\displaystyle\operatorname{Pr}(Y_j=1|Y_{j'}=0,X_j=0,X_{j'}=1)=q,\cr
\displaystyle \operatorname{Pr}(Y_j=1,Y_{j'}=1|X_j=1,X_{j'}=0)\cr
\displaystyle\quad{}\cdot\operatorname{Pr}(Y_j=0,Y_{j'}=0|X_j=1,X_{j'}=0)\cr
\displaystyle\quad{}/\bigl(\operatorname{Pr}(Y_j=0,Y_{j'}=1|X_j=1,X_{j'}=0)\cr
\displaystyle\hphantom{/}\quad {}\cdot\operatorname{Pr}(Y_j=1,Y_{j'}=0|X_j=1,X_{j'}=0)\bigr)=c.
}\hspace*{-12pt}
\end{equation}
It follows from results in \citet{Chen} that the joint distribution of
$Y_j$ and $Y_{j'}$ among the exposure--dis\-cordant pairs,
$\operatorname{Pr}(Y_j,Y_{j'}|X_j=1,X_{j'}=0)$, is completely defined
by the variation independent parameters $p$, $q$ and $c$. $c$
quantifies the degree of deviation from~(\ref{eq:factwins2}); in
particular, (\ref{eq:factwins2}) is violated when $c\neq 1$. It is easy
to show that assumption (\ref{eq:factwins1}) is logically compatible
with all joint values of ($p,q,c$). Thus, we proceed by assuming that
(\ref{eq:factwins1}) holds, so that the exposure--discordant crude
analysis consistently estimates $\psi_s=\psi_c$. Combining
(\ref{eq:ii}) and (\ref{eq:sim2}), and using results in \citet{Chen},
gives that $\hat{\psi}_{\mathit{c.clr}}$ converges to
\begin{equation}
\label{eq:conv} \log \biggl\{\frac{p(1-q)}{q(1-p)} \biggr\},
\end{equation}
whereas the true value of $\psi_c(=\psi_s=\psi_m)$ is given by
\begin{equation}
\label{eq:trueval} \log \biggl\{\frac{p(1-q)}{q(1-p)} \biggr\}+\log
\biggl\{\frac{1-q+qc}{1-p+pc} \biggr\}.
\end{equation}
Thus, the true value of $\psi_c$ depends on the association between
$Y_1$ and $Y_2$ through the second term in~(\ref{eq:trueval}), whereas
the asymptotic limit of $\hat{\psi}_{\mathit{c.clr}}$ does not. We used $p=0.3$,
$q=0.1$, and $c=4$.  For these values, $\psi_c=0.97$, whereas the
asymptotic limit of $\hat{\psi}_{\mathit{c.clr}}$ equals 1.35, for conditional
logistic regression. We generated 5000 samples, each consisting of
$n=2000$ exposure--discordant twin pairs. For each pair, the random
variables ($Y_1,Y_2$) were generated from the model in~(\ref{eq:sim2}).
Each sample was analyzed with conditional logistic regression (method~2), the mixed model (method~3) and the exposure--discordant crude
analysis (method~4). For these methods, we obtained an average estimate
of $\psi_c$ equal to 1.35, 1.26 and 0.97, respectively. Thus, both
conditional logistic regression and the mixed model produced biased
estimates, whereas the exposure--discordant crude analysis estimate was
unbiased.

Next, we consider misspecification of the function $b(Z;\gamma)$, in
the explicit model. We generated 5000 samples, each consisting of
$n=2000$ twin pairs. For each\vadjust{\goodbreak} twin pair, the random variables
$(Z,X_1,X_2,\break Y_1,Y_2)$ were generated from the model
\begin{equation}
\label{eq:sim5}
  \hspace*{12pt}\cases{
\displaystyle Z=(V,W),\cr
\displaystyle V\perp W,\cr
\displaystyle V\sim N(0,1),\cr
\displaystyle W\sim \operatorname{Ber}(0.5),\cr
\displaystyle X_1\perp X_2 | Z,\cr
\displaystyle \operatorname{logit}\{\operatorname{Pr}(X_j=1|Z)\}\cr
\displaystyle\quad=\alpha_0+\alpha_1V+\alpha_2W+\alpha_3VW,\cr
\displaystyle Y_1\perp Y_2 | (X_1,X_2,b),\cr
\displaystyle Y_j\perp X_{j'} | (X_j,b),\cr
\displaystyle \operatorname{logit}\{\operatorname{Pr}(Y_j=1|X_j,Z)\}=b(Z;\gamma)+\psi_cX_j,\cr
\displaystyle b(Z;\gamma)=\gamma_0+\gamma_1V+\gamma_2W+\gamma_3VW ,}\hspace*{-12pt}
\end{equation}
with $\alpha_0=2$, $\alpha_1=\alpha_2=1$, $\alpha_3=-1.5$,
$\gamma_0=-2$, $\gamma_1=\gamma_2=-1$, $\gamma_3=1.5$, $\psi_c=1.3$.
Each sample was analyzed with the misspecified explicit model\break
$\operatorname{logit}\{\operatorname{Pr}(Y_j=1|X_j,Z)\}=\gamma_0+\gamma_1V+\gamma_2W+\psi_cX_j$.
We obtained an average estimate of $\psi_c$ equal to 0.69, which is
severly biased.

Finally, we consider the assumption that the random effect
$b(Z;\gamma)$ is normally distributed, which is commonly made for mixed
models. Toward this end we reanalyzed the 5000 samples generated from
model (\ref{eq:sim5}), now fitting the mixed model
$\operatorname{Pr}(Y=1|X,b)=b+\psi_cX$ to the exposure--discordant
pairs, with $b|X_1\neq X_2 \sim N(\theta,\sigma^2)$. Under the data
generating model, the conditional distribution of $b(Z;\gamma)$, given
$X_1\neq X_2$ is rather complicated, and, in particular, not normal. We
obtained an average estimate of $\psi_c$ equal to 1.30, which is
identical to the true value, to the second decimal. This finding
supports the theoretical results in \citet{Neuhaus2}, which state that
the mixed model is robust against the normal random effect assumption.

\section{Real Data Examples}
\label{sec:examples}

\subsection{Matched Cohort Data}

The first example is taken from a matched cohort study that aimed to
investigate the effect of hysterectomy on risk for CVD \citep{Ingelsson}. A common surgery among
perimenopausal women, hysterectomy is often performed on benign
indications, but its long-term consequences are not fully
understood. The study is  based on the Swedish Inpatient Register,
where all women who underwent hysterectomy between January 1973 and
December 2003 (227,389 individuals) were identified. For each
hysterectomized woman, three women who never had hysterectomy were
randomly selected from the Register of Total Population. The three
unexposed wom\-en were individually matched to the exposed woman by
birth year, year of hysterectomy, and county of residence at year of hysterectomy.

Information on CVD status was obtained from the Inpatient Register
and information of follow up through record linkage to the Cause of
Death Register, Emigration Register and Cancer Register. To avoid
bias from CVD events occurring in relation to the hysterectomy
surgery, the exposed women started their risk time from 30 days
after hysterectomy; they were then followed until CVD, heart
failure, cervical, corpus or ovarian cancer, death, emigration or
end of study (Dec 31, 2003). Similarly, unexposed women started
their risk time 30 days after the date of matching, that is, the date of
hysterectomy of the corresponding exposed woman. For further details
on the study, see \citet{Ingelsson}.

In the current analysis we focus on 1:1 matched studies with binary
outcomes. We constructed a binary outcome by defining $Y=1$ for women
who developed CVD during follow-up, and $Y=0$ for the remaining women.
We constructed a 1:1 matched sample by matching each exposed woman to
one unexposed woman, which was randomly selected from the three
unexposed women in the same set. After the exclusions described above,
we ended up with 52,814 1:1 matched pairs, of which 6712 were
discordant in both the exposure and the outcome. The data were analyzed
with methods 1--4 described in Section~\ref{sec:sim}. For method 1 we
used the explicit model
$\operatorname{logit}\{\operatorname{Pr}(Y=1|Z,X)\}=\gamma_0+\gamma_1[\mbox{birth
year}]+\gamma_2[\mbox{year at
hysterectomy}]+\gamma_3[\mbox{county}]+\psi_cX$,\break where $\gamma_3$ is a
factor parameter with one level for each county.

\begin{table}[b]
\tabcolsep=0pt \caption{Analysis results for the 1:1 matched subset of
the hysterectomy-CVD data} \label{tab:tab4}
\begin{tabular*}{\columnwidth}{@{\extracolsep{\fill}}lccc@{}}
\hline
\textbf{Analysis method} & \textbf{Target parameter} & \textbf{Point est} & \textbf{95\% CI} \\
\hline
1. Explicit & $\psi_c$ & 0.03  & $-$0.02, 0.08\\
2. Cond log reg & $\psi_c$ & 0.03  & $-$0.02, 0.08\\
3. Mixed discordant & $\psi_c$ &  0.03  & $-$0.02, 0.08\\
4. Crude discordant & $\psi_s$ & 0.03 & $-$0.02, 0.07\\
\hline
\end{tabular*}
\end{table}

Table \ref{tab:tab4} displays the results. For all three methods, there
is a significant (at 5\% level) association between hysterectomy and
CVD. The point estimates obtained by conditional logistic regression
and expo\-sure--discordant crude analysis are almost identical, whereas
the point estimate obtained from the mixed model is twice as large.
According to theory (Neuhaus et al., \citeyear{Neuhaus2}) we would expect the mixed model
estimate to be identical\vadjust{\goodbreak} to the estimate obtained from conditional
logistic regression. Indeed, methods 1--4 all give identical estimates
to the second decimal.\looseness=1

Although our focus is on 1:1 matching,  all methods in this paper
generalize directly to $m$:$n$ matching (see Section~\ref{sec:discussion}). Table \ref{tab:tab8} displays the results when
the whole 1:3 matched data is analyzed, using methods 1--4 described in
Section~\ref{sec:sim}.

\begin{table}
\tabcolsep=0pt
\caption{Analysis results for the full 1:3 matched hysterectomy-CVD~data} \label{tab:tab8}
\begin{tabular*}{\columnwidth}{@{\extracolsep{\fill}}lccc@{}}
\hline
\textbf{Analysis method} & \textbf{Target parameter} & \textbf{Point est} & \textbf{95\% CI} \\
\hline
1. Explicit & $\psi_c$ & 0.06 & 0.02, 0.09\\
2. Cond log reg & $\psi_c$ & 0.06 & 0.02, 0.09\\
3. Mixed discordant & $\psi_c$ &  0.06  & 0.02, 0.09\\
4. Crude discordant & $\psi_s$ & 0.05  & 0.02, 0.09\\
\hline
\end{tabular*}   \vspace*{3pt}
\end{table}

\subsection{Twin Data}
The second example is from a twin study of the association between
fetal growth and asthma \citep{Ortqvist}. Several studies have shown
that there is an association between asthma and low birth weight.
This association could potentially be explained by a~causal effect
of impaired fetal growth on asthma, but may also be explained by
confounding factors. In particular, gestational age is correlated
with both birth weight and asthma, and may confound the birth
weight-asthma association ({\"O}rtqvist et~al.,\break \citeyear{Ortqvist}). Twins provide an
excellent opportunity to separate the causal effect of birth weight
from the confounding effect of gestational age, and at the same time
adjust for other shared familial factors.

All twins born in Sweden in June 1992 to June 1998 were identified
through the Swedish Twin Register at the age of 9 or 12 years.
Information on asthma and zygosity was collected in telephone
interviews with their parents. Birth weight was retrieved from the
Medical Birth Register (MFR). Of the 15,808 eligible twins 69\%
(10,918 individuals) had information on asthma and could also be
securely linked to the MFR. In total, there were 3107 MZ pairs. 1087 pairs were discordant in birth weight (exposure), where discordance was defined as a difference
greater than 400 grams or 15\%, and 175 pairs were discordant on
both birth weight and asthma (outcome).

The data were analyzed using methods 2--6 described in Section~\ref{sec:sim}. Table \ref{tab:tab3} displays the results. The estimates
obtained from conditional logistic\vadjust{\goodbreak} regression and the
exposure--discordant crude analysis are both smaller than estimate
obtained from the all-pair crude analysis. This finding suggests that
the birth weight-asthma association is inflated by shared confounding.
Methods 2, 3 and 6 gave very similar results, as predicted by theory
(Neuhaus et~al., \citeyear{Neuhaus2}; \citecs{Neuhaus}).

\begin{table}
\tabcolsep=0pt
\caption{Analysis results for the birth weight-asthma twin data}
\label{tab:tab3}
\begin{tabular*}{\columnwidth}{@{\extracolsep{\fill}}lccc@{}}
\hline
\textbf{Analysis method} & \textbf{Target parameter} & \textbf{Point est} & \textbf{95\% CI} \\
\hline
2. Cond log reg & $\psi_c$ & 0.29 & $-$0.01, 0.59  \\
3. Mixed discordant &  $\psi_c$ & 0.29 & $-$0.01, 0.59  \\
4. Crude discordant & $\psi_s$ &  0.18 & $-$0.01, 0.37 \\
5. Crude all & $\psi_m$ & 0.33 & \phantom{$-$}0.16, 0.50 \\
6. Mixed all & $\psi_c$ &  0.30 &  \phantom{$-$}0.00, 0.60 \\
\hline
\end{tabular*}  \vspace*{3pt}
\end{table}

\section{Discussion}
\label{sec:discussion}

We have given an overview of the most common analysis methods for
matched cohort studies. We have identified the target parameters in
each method, outlined the underlying assumptions and compared the
methods in terms of statistical power. The analysis methods that we
have considered do not estimate the same parameter; the exposure--discordant crude analysis and the within--between model estimate a
standardized odds ratio, whereas the explicit method, conditional
logistic regression, and the mixed model \mbox{estimate} a conditional odds
ratio. Thus, the choice between these methods should primarily be
guided by the research question being asked. In addition, it is also
important to consider the statistical power, underlying assumptions,
computer intensity and flexibility of the methods. Theoretical
arguments suggest that when all underlying assumptions hold, all
methods that we have considered have the same statistical power. This
was confirmed in our simulation study. In terms of underlying
assumptions, the methods differ significantly. The exposure--discordant
crude analysis relies on few\-er assumptions than the other methods. In
terms of computer intensity, the mixed model requires numerical
optimization, and is far more time consuming than the other methods. In
terms of flexibility, all methods, except the expo\-sure--discordant
crude analysis, most naturally target odds ratios. The
expo\-sure--discordant crude analysis however, can easily be used to
target any measure of the exposure-out\-come association.

We have considered 1:1 matching. Frequently, $m$:$n$ matching is
employed, that is, each set is constructed by matching $m$ exposed
subjects to $n$ unexposed subjects. All methods in this paper
generalize directly to $m$:$n$ matching. Specifically, the underlying
assumptions and the interpretation of the target parameters remains the
same under $m$:$n$ matching. We conjecture that many of the theoretical
properties that we have derived for 1:1 matching carry over to $m$:$n$
matching as well, for example, the asymptotic equivalence in terms of
power. However, a stringent treatment of $m$:$n$ matching is more
difficult. For instance, under violation of (\ref{eq:const}) the
probability limit of $\hat{\psi}_{\mathit{c.clr}}$ has no longer an analytic
expression, which hampers a theoretical comparison with the
exposure--discordant crude analysis. Comparing the methods under
$m$:$n$ is a topic for future research.\looseness=1

In practice, it is often desirable to adjust the analysis for
additional covariates which are not matched on. In the model-based
methods (i.e., all methods except the exposure--discordant crude
analysis), adjustment for additional covariates can easily be
accomplished by adding the covariates as a regressor in the model. It
is not obvious though, how to adjust for additional covariates in the
exposure--discordant crude analysis. Extensions of the
exposure--discordant crude analysis for additional covariate
adjustments is a topic for future research.

\begin{appendix}

\def\theequation{\arabic{equation}}

\section{}
\label{sec:app2} \setcounter{equation}{31}

 Define
$p^x=\operatorname{Pr}(Y\!=\!1|X\!=\!x)$, $q=\operatorname{Pr}(X\!=\!1)$,
$q^{00}=\operatorname{Pr}(X_1\!=\!X_2\!=\!0)$,
$q^{11}=\operatorname{Pr}(X_1\!=\!X_2\!=\!1)$,
$q^d=\operatorname{Pr}(X_1\!\neq\!
X_2)$, $c^{00}=\operatorname{cov}(Y_1,Y_2|X_1\!=\!X_2\!=\!0)$,
$c^{11}=\operatorname{cov}(Y_1,Y_2|\break X_1\!=\!X_2\!=\!1)$,
$c^d=\operatorname{cov}(Y_1, Y_2|X_1\!\neq\! X_2)$,
$\psi_0=\operatorname{logit}(p_0)$,
$\psi_m=\operatorname{logit}(p_1)- \operatorname{logit}(p^0)$ and
$\psi=(\psi_0,\psi_m)^T$. $\hat{\psi}_{m.crude}$ in (\ref{eq:ty}) can
be expressed as the second element of the solution to
$\sum_iU_i(\psi)=0$, where
\begin{eqnarray*}
&&U_i(\psi)\\
&&\quad=  \left\{\matrix{
(1-X_{i1})(Y_{i1}-p^0)+(1-X_{i2})(Y_{i2}-p^0)\cr
X_{i1}(Y_{i1}-p^1)+X_{i2}(Y_{i2}-p^1)} \right\}.
\end{eqnarray*}
It follows from standard theory that $n^{1/2}(\hat{\psi}-\psi)$ is asympotically normal with mean 0 and variance
\[\label{eq:var}
  \biggl[ E \biggl\{\frac{\partial U_i(\psi)}{\partial\psi^T} \biggr\}  \biggr] ^{-1}\operatorname{var}\{U_i(\psi)\}  \biggl[   \biggl[ E \biggl\{\frac{\partial U_i(\psi)}{\partial\psi^T} \biggr\}  \biggr] ^{-1}  \biggr] ^T,
\]
where, after some algebra,
\[
E \biggl\{\frac{\partial U_i(\psi)}{\partial\psi^T} \biggr\}= \pmatrix{
-2p^0(1-p^0) & 0\cr -2p^1(1-p^1) & -2p^1(1-p^1)}
\]
and{\fontsize{9}{11}{\selectfont{
\begin{eqnarray*}
&&\hspace*{-5pt}\operatorname{var}\{U_i(\psi)\}\\
&&\hspace*{-5pt}\quad=  \pmatrix{
2(1-q)p^0(1-p^0)+q^{00}c^{00} & q^dc^d\cr q^dc^d &
2qp^1(1-p^1)+q^{11}c^{11} }.
\end{eqnarray*}}}}%
After additional algebra, the asymptotic variance for $n^{1/2}(\hat{\psi}_{m.crude}-\psi_m)$ is obtained as
\begin{eqnarray}
\label{eq:vm}
&&\frac{1}{2(1-q)p^0(1-p^0)}+\frac{1}{2qp^1(1-p^1)}\nonumber\\
&& \quad{}+\frac{q^{00}c^{00}}{4\{p^0(1-p^0)\}^2}
+\frac{q^{11}c^{11}}{4\{p^1(1-p^1)\}^2}\\
&&\quad{}-\frac{q^dc^d}{2q(1-q)p^0(1-p^0)p^1(1-p^1)}.\nonumber
\end{eqnarray}
Replacing the population parameters in (\ref{eq:vm}) with their sample
counterparts gives the standard error for $\hat{\psi}_{m.crude}$.

To derive the standard error formula in (\ref{eq:se}) we note that a
regular 1:1 matched cohort can be obtained by setting $q=0.5$,
$q^{00}=q^{11}=0$ and $q^d=1$. The expression in (\ref{eq:vm}) then
simplifies to
\begin{eqnarray}
\label{eq:v} &&\frac{1}{p^0(1-p^0)}+\frac{1}{p^1(1-p^1)}\nonumber
\\[-8pt]\\[-8pt]
&&\quad{}-\frac{2c^d}{p^0(1-p^0)p^1(1-p^1)}.\nonumber
\end{eqnarray}
Replacing the population parameters in (\ref{eq:v}) with their sample
counterparts gives the standard error formula in~(\ref{eq:se}).

\section{}
\label{sec:app4}

Define $\psi_c^{\dagger}=\log  \{
\frac{\operatorname{Pr}^*(Y^1=1,Y^0=0)}{\operatorname{Pr}^*(Y^0=1,Y^1=0)}
\}$, $\mathrm{H}_c^{\dagger}\dvtx  \psi_c^{\dagger}=0$,
$\psi_s^{\dagger}=\log
\{\frac{\operatorname{Pr}^*(Y^1=1)\operatorname{Pr}^*(Y^0=0)}{\operatorname{Pr}^*(Y^1=0)\operatorname{Pr}^*(Y^0=1)}
\}$, $\mathrm{H}_s^{\dagger}\dvtx \psi_s^{\dagger}=0$.
$\mathrm{H}_c^{\dagger}$ can be tested using the likelihood ratio test
(LRT) statistic
\[
T_{c,LR}^{\dagger}=-2\log \biggl \{\frac{ \sup
_{\mathrm{H}_c^{\dagger}}(p_{00}^{W}p_{01}^Up_{10}^Vp_{11}^T)}{\sup(p_{00}^{W}p_{01}^Up_{10}^Vp_{11}^T)}
\biggr\},
\]
and $\mathrm{H}_s^{\dagger}$ can be tested using the LRT statistic
\[
T_{s,LR}^{\dagger}=-2\log  \biggl\{\frac{ \sup
_{\mathrm{H}_s^{\dagger}}(p_{00}^{W}p_{01}^Up_{10}^Vp_{11}^T)}{\sup(p_{00}^{W}p_{01}^Up_{10}^Vp_{11}^T)}
\biggr\},
\]
where $p_{y^0y^1}\hspace*{-0.5pt}=\hspace*{-0.5pt}\operatorname{Pr}^*(Y^0\hspace*{-0.5pt}=\hspace*{-0.5pt}y^0,Y^1\hspace*{-0.5pt}=\hspace*{-0.5pt}y^1)$, and the
suprema are taken under the restrictions $0<p_{y^0y^1}<1$  and $\sum
_{y^0y^1}p_{y^0y^1}=1$. Regardless of whether (\ref{eq:const}),
(\ref{eq:faca}) and (\ref{eq:facb}) hold or not, $\hat{\psi}_{\mathit{c.clr}}$
and $\hat{\psi}_{\mathit{s.crude}}$ are the nonparametric MLEs of
$\psi_c^{\dagger}$ and $\psi_s^{\dagger}$, respectively. Thus,
$T_{c,LR}^{\dagger}$ and $T_c$ are asymptotically equal, and
$T_{s,LR}^{\dagger}$ and\vadjust{\goodbreak} $T_s$ are asymptotically equal. It is easy to
show that $\mathrm{H}_c^{\dagger}$ and $\mathrm{H}_s^{\dagger}$ are
equivalent (i.e., $\mathrm{H}_c^{\dagger}$ holds if and only
if~$\mathrm{H}_s^{\dagger}$ holds), which implies that
$T_{c,LR}^{\dagger}$ and $T_{s,LR}^{\dagger}$ are identical, which then
in turn implies that $T_c$ and $T_s$ are asymptotically equal.

It is easy to show that $\mathrm{H}_0$ and (\ref{eq:faca}) together
imply~$\mathrm{H}_s^{\dagger}$, and thus also $\mathrm{H}_c^{\dagger}$.
Because $\hat{\psi}_{\mathit{c.clr}}$ converges to~$\psi_c^{\dagger}$, it then
follows that $\hat{\psi}_{\mathit{c.clr}}$ converges to 0 under~$\mathrm{H}_0$
and~(\ref{eq:faca}).

\section{}
\label{sec:app1}

Under (\ref{eq:sim}), we have that
\begin{eqnarray*}
&&\operatorname{Pr}(X_1,X_2,b)
\\
&&\quad=\frac{1}{\sqrt{2\pi}}e^{ \{b-\theta\bar{X} \}^2/2}\operatorname{Pr}(X_1,X_2)\\
&&\quad=h(X_1,b)h(X_2,b)e^{-\theta^2X_1X_2/4}\operatorname{Pr}(X_1,X_2),
\end{eqnarray*}
for some function $h(\cdot,\cdot)$. $X_1\perp X_2|b$ now implies that
\[
e^{-\theta^2X_1X_2/4}\operatorname{Pr}(X_1,X_2)=k(X_1)k(X_2)
\]
for some function $k(\cdot)$, which in turn implies that
$\theta=2\sqrt{\log(\phi)}$.
\end{appendix}

\section*{Acknowledgment}
Arvid Sj\"olander acknowledges financial support from The Swedish Research Council (2008-5375).


\end{document}